\documentclass[12pt]{article}
\usepackage{graphics}
\usepackage{epsf}
\usepackage{amsfonts}
\usepackage{amssymb}

\catcode `@=11 \@addtoreset{equation}{section} \catcode `@=12

\def\tsector#1#2{\ {\scriptstyle #1}\hskip 1mm
\mathop{\opensquare}\limits_{\lower 1mm\hbox{$\scriptstyle#2$}}^\sim\hskip 1mm}
\def\appendix{{\newpage\section*{Appendix}}\let\appendix\section%
        {\setcounter{section}{0}
        \gdef\thesection{\Alph{section}}}\section}

\def\){\right)}
\def\({\left( }
\def\]{\right] }
\def\[{\left[ }

\newcommand{\be}{\begin{equation}}
\newcommand{\ee}{\end{equation}}
\newcommand{\ba}{\begin{eqnarray}}
\newcommand{\ea}{\end{eqnarray}}
\newcommand{\no}{\nonumber \\}

\def\C{{\mathbb{C}}}
\def\Z{{\mathbb{Z}}}
\begin{document}
\title {{\small \hfill SLAC-PUB-10091 }~~~~~~~\\
 Comments on the Fate of unstable orbifolds}
\author{ Sang-Jin Sin \\   \small \sl
Stanford Linear Accelerator Center, Stanford University, Stanford
CA 94305 \footnote{Work supported partially by the department of Energy under contract number DE-AC03-76SF005515. }
\\ \small \sl and \\
\small  \sl 
Department of Physics, Hanyang University, Seoul, 133-791, Korea \footnote{Permanent address} 
} 

\maketitle
\begin{abstract}
We study the localized tachyon condensation in their mirror Landau-Ginzburg picture.
We completely determine the decay mode of an unstable orbifold  $C^r/Z_n$, $r=1,2,3$ under 
the condensation of a tachyon with  definite R-charge and mass by  extending  the Vafa's work hep-th/0111105.
Here, we give a simple method  that works uniformly for all $C^r/Z_n$. 
For $C^2/Z_n$,  where method of toric geometry works, we give  a proof of equivalence of our method  with  toric one. 
For $C^r/Z_n$ cases, the orbifolds decay into  sum of $r$ far separated orbifolds. 
\end{abstract}

\newpage
\section{Introduction}
The study of open string tachyon condensation\cite{sen} has led to many interesting consequences 
including classification of the D-brane charge by K-theory. 
While the closed string tachyon condensation involve the change of the background spacetime 
and much more difficult, if we consider the case where tachyons can be localized 
at the singularity,  one may expect the maximal analogy with the open string case.
Along this direction, the study of localized tachyon  condensation  was  considered in \cite{aps}
using the brane probe and renormalization group flow and by many others\cite{vafa,hkmm,dv,dab,sarkar,many}. 
The basic picture is that 
tachyon condensation induces cascade of decays of the orbifolds to  less singular ones until the spacetime 
supersymmetry is restored. Therefore the localized tachyon condensation has geometric description as the resolution 
of the  spacetime singularities. 

Soon after, Vafa\cite{vafa} considered the problem in the Landau-Ginzburg (LG) formulation using the 
Mirror symmetry and confirmed the result of \cite{aps}.
In \cite{hkmm}, the same problem  is studied by using the RG flow as deformation 
of chiral ring and in term of toric geometry.
In \cite{vafa}, Vafa showed that, as a consequence of the tachyon
condensation, the final point of the  process is sum
of two orbifold theories which are far from each other but smoothly connected: 
one located at north and the other at
the south poles of blown up $P^2$ singularity of the orbifold in
the limit where the radius of the sphere is infinite.
Schematically, we can represent this transition by 
\be
\C^2/\Z_{n(k_1,k_2)} \to \C^2/\Z_{p_1(*,*)} \oplus \C^2/\Z_{p_2(*,*)}, \label{unk}\ee 
with yet unknown generators for the daughter theories. 

The purpose of this paper is to determine the  the decay mode of unstable orbifolds by 
working out the generators of orbifold action in daughter theories for  $C^r/Z_n$ $r=1,2,3$. 
For $\C^1/\Z_{n}$, the transition modes are  described in earlier works \cite{aps,vafa,hkmm}.
For $\C^2/\Z_{n(k_1,k_2)}$ case, some examples are worked out in \cite{hkmm} using toric geometry and prescription
in terms of continued fraction is given. In principle, it can be worked out once numbers are given explicitly. 
However, that method does not work for $C^3/Z_n$.
Here, we give a simple method  that works easily and uniformly for all $C^r/Z_n$.
For $C^2/Z_n$, we give  a proof of equivalence of our method  with  toric one. 
To do this we will need to know how the spectrums of chiral primaries are transformed under the
condensation of a specific tachyon.  

\section{Mirror symmetry and Orbifolds}
We begin by a summary of Vafa's work \cite{vafa} on  localized  tachyon condensation. 
The orbifold $\C^r/Z_n$ is defined by the  $\Z_n$ action given by equivalence relation 
\be
(X_1,...,X_r)\sim(\omega^{k_1}X_1,...,\omega^{k_r}X_r),\quad
\omega=e^{2\pi i /n} .\label{znaction}\ee 
We call $(k_1,\cdots, k_r)$ as the generator of the $\Z_n$ action.
The orbifold can be imbedded into the gauged linear sigma model(GLSM) \cite{wittenN2}.
The vacuum manifold of the latter  is described by the D-term constraints 
\be -n|X_0|^2+\sum_i k_i |X_i|^2=t .\ee 
Its $t\to -\infty$ limit corresponds to the orbifold and 
the $t\to\infty$ limit is the
$O(-n)$ bundle over the weighted projected space 
$WP_{k_1,...,k_r}$.
$X_0$ direction corresponds to the non-compact fiber of this bundle and $t$ plays role of 
size of the $WP_{k_1,...,k_r}$.

By dualizing  this GLSM, we get a LG model with  a superpotential\cite{HV}
\be W=\sum_{i=0}^r \exp(-Y_i), \ee
 where twisted chiral fields
$Y_i$ are periodic $Y_i\sim Y_i+2\pi i$ and  related to $X_i$ by
$Re[Y_i]=|X_i|^2.$ Introducing  the variable $ u_i:=e^{-Y_i/n},$ the D-term  constraint is expressed as 
$e^{-Y_0}=e^{t/n}\prod_i
u^{k_i} $. 
The periodicity of $Y_i$ imposes the identification : 
$u_i \sim e^{2\pi i/n} u_i $ which necessitate  modding out each $u_i$ by $\Z_n$.  
The result  is usually described by  
\be
[W=\sum_{i=1}^r u_i^{n}+e^{t/n}\prod_i u^{k_i} ]// (\Z_n)^{r-1}.\label{orLGeq}
\ee
which describe  the  mirror Landau-Ginzburg model of the linear sigma model. 
As a $t\to -\infty$ limit, mirror of the orbifold is  
\be
[W=\sum_{i=1}^r u_i^{n} ]// (\Z_n)^{r-1}. 
\ee 
 Since it is  not ordinary Landau-Ginzburg theory but an orbifolded  version,
the chiral ring structure of the theory is very different from
that of LG model. For example, the dimension of the local ring of
the super potential is always $n-1$, regardless of $r$.
  
We list some properties of orbifolded LG theory for later use.

The true variable of the theory are $Y_i$ not $u_i$ related by $u_i=e^{-Y_i/n}$.
As a consequence,  monomial basis of the chiral ring is given by
\be
\{u_1^{p_1}u_2^{p_2}|(p_1,p_2)=(n\{jk_1/n\},n\{jk_2/n\}),
j=1,...,n-1\} \label{basis},
\ee
and $u_1^{p_1}u_2^{p_2}$ has weight $(p_1,p_2)$ and charge $(p_1/n,p_2/n)$.

\section{Fate of the spectrum} 

For $\C^2/\Z_{n(k_1,k_2)}$ case, if one consider the
condensation  of tachyon in the $l$-th twisted sector that  corresponds 
to chiral ring element 
$u_1^{p_1}u_2^{p_2}$, with $p_1=n\{lk_1/n\}$ and
$p_2=n\{lk_2/n\}$, the theory is given by the super
potential \be [W=u_1^n+u_2^n +e^{t/n}u_1^{p_1}u_2^{p_2}]//Z_n .
\ee 

Consider $u_2 \sim 0$ and $u_2^n \sim e^{t/n}u_1^{p_1}u_2^{p_2}$ 
region, which should be described by  
\be [W\sim
u_1^n+e^{t/n}u_1^{p_1}u_2^{p_2}]//Z_n.\ee 
By introducing the new variables $v_1=u_1^{n/p_2}$ and $v_2=e^{t/np_2}u_1^{p_1/p_2}u_2$.
The single valuedness of $v_i$ induces the $Z_n$ but single
valuedness of $u_1^n$ and $u_1^{p_1}u_2^{p_2}$ implies that
$v_1,v_2$ are orbifolded by $Z_{p_2}$.  By substitution, we can
express  $u_1^{q_1}u_2^{q_2}$ in terms of $v_1,v_2$: \be
u_1^{q_1}u_2^{q_2}= v_1^{Q_1}v_2^{Q_2},\ee {\rm where} \be
(Q_1,Q_2)=(-p\times q/n, q_2),\ee with $p\times
q=(p_1q_2-p_2q_1)$. Notice that  map  $T^-_p: (q_1,q_2)\mapsto
(Q_1,Q_2)$  is linear map acting on the integrally normalized weight space 
and can be described by a matrix
\be T^-_p=\pmatrix{{p_2}/{n} & -{p_1}/{n} \cr 0 & 1}. \label{Tp-}
\ee  It is working near $u_2\sim 0$. It
maps $(n,0) \to (p_2,0)$ and $(p_1,p_2) \to (0,p_2)$, or equivalently, 
$u_1^n\to v_1^{p_2}$ and $u_1^{p_1}u_2^{p_2}\to
v_2^{p_2}$.  

One should notice that $Q_1,Q_2$ are not integers in general.
However, when both $p$ and $q$ are weight vectors of elements of
orbifold chiral ring, generated by $(k_1,k_2)$,  they are
integers. This is because if $p=(n\{lk_1/n\},n\{lk_2/n\}),
q=(n\{jk_1/n\},n\{jk_1/n\})$, $s:=p\times q/n$, then
 \be s=n\{lk_1/n\}\{jk_2/n\}-n\{lk_2/n\}\{jk_1/n\}\in \mathbb{Z} \label{4.7} \ee
 for any integers $n,k,l,j$.  For $k_1=1$, $s=-l[jk_2/n]+j[lk_2/n]$.
Especially interesting case will be $q=k=(1,k_2)$, in which case,
we have $s=[lk_2/n]=(lk_2-p_2)/n$. Geometrically, $s$ is
proportional to the area spanned by two vectors $p$ and $q$.
Therefore it is zero if $p$ and $q$ are parallel.

The R-charges
are determined by the marginality condition. In the original
theory, $u_i$ has R-charge $1/n$ since $u_i^n$ has R-charge 1. We express this as $R[u_i^n]=1$. 
Therefore
$R[u_1^{p_1}u_2^{p_2}]=(p_1+p_2)/n$. 
charge space is defined by the weight space scaled by $1/n$. So we use the same figure 1 to describe it.
The diagonal in charge space
is the line connecting A$(1,0)$ and B$(0,1)$. Any operator whose R
charge is on this diagonal corresponds to the marginal operator.
The points below the diagonal correspond to the relevant operators and tachyonic
and those above it correspond to the irrelevant operators.  
 When  a tachyon, P, is fully condensed, the marginal line is changed from diagonal line AB to line AP or BP.
 AP gives down-theory and BP gives the up-theory.
$\Delta_+$ is the cone spanned by $\vec{OB}$ and $\vec{OP}$, and
similarly $\Delta_-$ is the cone spanned by $\vec{OA}$ and $\vec{OP}$.

\begin{figure}[htbp2]
\epsfxsize=7cm
    \centering
\centerline{\epsfbox{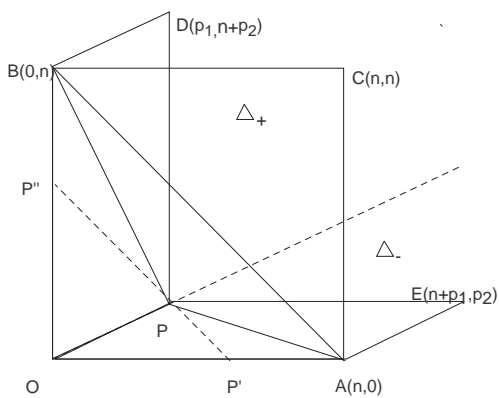}}
    \caption{\scriptsize 
    Integrally normalized weight/charge space for $\C^2/\Z_n$. It can be considered as the space of power of local ring elements. 
 It is defined as a two dimensional torus with size n.   $u_1^n$ and $u_2^n$ is  located at A(n,0) and B(0,n) respectively.  
 Under the  condensation of tachyon P, the parallelogram $OBDP$ is mapped to the
 up-theory and $OPEA$ is mapped to the down-theory. Translation parallel to 
 OP is mapped to horizontal in up theory and vertical in down theory. }
\label{Fig2}
\end{figure}

Let P be the point $(p_1/n,p_2/n)$ in charge space that
corresponds to a chiral primary that is undergoing condensation, and
Q be any charge point $(q_1/n,q_2/n)$ and A, B now corresponds to
$(1,0)$ and $(0,1)$. 
One can work out the action of $T^-_p$ from other point of view.
If P represent the chiral primary of $l$-th
twisted sector, $(p_1/n,p_2/n):=(\{lk_1/n\},\{lk_1/n\})$. Near
$u_2\sim 0$ region, the marginality condition is changed to
$R[u_1^{p_1}u_2^{p_2}]=1, R[u_1^n]=1$. In terms of new variable
$R[v_i^{p_2}]=1$. The linear transformation 
\be {\tilde T_p^-}: (q_1/n,q_2/n) \to (Q_1/p_2,Q_2/p_2),\ee
can be determined by its action on P and (1,0). 
Once ${\tilde T_p^-}$ is decided, we get ${T_p^-}$ from the relation, 
${\tilde T_p^-}=\frac{n}{p_2}T_p^-$.  The result of course agrees with the one given by eq.(\ref{Tp-}).
Under this mapping,  the lower triangle $ \triangle POA$ in figure {\ref{Fig2}}
in charge space is mapped to the entire $\triangle BOA$, which
defines one of theory in the final stage of the tachyon
condensation. We call it down-theory. 
\footnote{Conversely, if we require that ${\tilde T_p^-}$
maps $\triangle POA$ to $\triangle BOA$,  ${\tilde T_p^-}$ is completely determined. 
The mapping $T^-$ in the integrally normalized weight space is induced  by
$T^-=(p_2/n){\tilde T^-}$. The normalization is dictated from the
condition that $T$ maps from  integer vectors to integer vectors.
Finally $T^-_p(n,0)=(p_2,0)$ and $T^-_p(p_1,p_2)=(0,p_2)$ so that
the identification $u_1^n=v^{p_2},\; u_1^{p_1}u_2^{p_2}=v^{p_2}$
is dictated.}

Similarly, by considering  $u_1\sim 0$ region, we get the mapping
${\tilde T}_p^+$ that maps the upper triangle $\triangle BOP$ to
$\triangle BOA$. By the relation $T^+_p=(p_1/n){\tilde T}^+_p$ we
can obtain the mapping in weight space:
 \be 
 T_p^+ q= \left(
\begin{array}{cc} 1 & 0\\
- p_2/n & {p_1}/{n}
\end{array}\right){q_1\choose q_2}= {q_1\choose p\times q/n}
\ee 
Notice that $T_p^+$ leaves all the vertical lines in weight
space fixed while $T_p^-$ leaves horizontal lines  invariant. 
 
Now we ask: given an operator with $q=(q_1,q_2)$, should we map
with $T^+_p$ or $T^-_p$? The answer is that we should use
the map that gives smaller R-charge. The difference of the
R-charge after the mapping is given by 
\ba
\delta:=R[T^+_pq]-R[T^-_pq]=\frac{p\times q}{np_1p_2}(p_1+p_2-n)
&<0& \;{\rm if}\; q \in \Delta_+ \no &> 0& \;{\rm if}\; q \in
\Delta_- ,\ea 
where $\Delta_+$ is the cone spanned by $\vec{OB}$
and $\vec{OP}$, and  similarly $\Delta_-$ is the cone spanned by
$\vec{OA}$ and $\vec{OP}$.  Notice that we are condensing relevant
operator $p$ so that $p_1+p_2<n$. The line $BP$ is mapped to the
marginal line of a final theory, the up-theory, and the line $AP$
is mapped to that of down-theory. Therefore the  emerging picture
is following: The parallelogram $OBDP$ spanned by $\vec{OB}$ and
$\vec{OP}$ is mapped to the up-theory whose weight space size is
$p_1$. Similarly, the parallelogram $OPEA$ spanned by $\vec{OP}$
and $\vec{OA}$ is mapped to the down-theory whose weight space
size is $p_2$. See figure 2. 
From eq. (\ref{4.7}), it is easy to see that  chiral ring elements 
of Mother theory are mapped to chiral ring elements of the daughter theories, 
under the condensation of a chiral ring element. Any operator $q'$ outside these two
parallelograms can be parallel translated to inside one of above
two parallelograms by the vector $\vec{OP}$ a few times if
necessary.  In daughter theories, if $q'\in \Delta_+$, then $T_p^+q'$ can
be translated horizontally by $p_1$ a few times to a point in the
up-theory. Similarly,  if $q'\in \Delta_-$, then $T_p^-q'$ can be
translated vertically by $p_1$ a few times to a point in the
up-theory.

\section{Fate of unstable orbifolds}
\subsection{ $\C^2/\Z_n$}
We now can answer to our main question: what are the generators of final theories? 
We noticed that there are two theories in the final stage. These two theories are
described by the difference of the marginal lines in the weight
space: extension of $BP$ or that of $AP$. We call the
former as the up-theory, describing $u_1 \sim 0$ region, and the
latter as down-theory, describing the $u_2\sim 0$ region. In terms
of the charge space, up-theory is obtained by mapping ${\tilde
T}^+_p: \triangle BOP \mapsto \triangle BOA$ and down-theory is
obtained by mapping ${\tilde T}^-_p: \triangle BOP \mapsto
\triangle BOA$.

The up-theory is a orbifold   $\C^2/\Z_{p_1}$ and the down theory
is another orbifold $\C^2/\Z_{p_2}$. Let $k=(k_1,k_2)$ be the
generator of the original theory. Then the generator of the
up-theory is given by $T_p^+(k)=(k_1, p\times k/n)$ and that of
the $T_p^-(k)=(-p\times k/n,k_2)$. Since $(k_1,k_2) \sim (-k_1,-k_2)$ as a
generator, one can also use
$T_p^-(-k)=(p\times k/n,-k_2)$ instead of $T_p^-(k)$. 
 Therefore we can describe the
process of condensation of tachyon with charge $p=(p_1,p_2)$ as
follows: \be \C^2/\Z_{n(k_1,k_2)} \longrightarrow
\C^2/\Z_{p_1(k_1,p\times k/n)} \oplus \C^2/\Z_{p_2(-p\times
k/n,k_2)} .\ee
 To simplify the notation, we use $n(k_1,k_2)$ for $\C^2/\Z_{k_1,k_2}$
and $s=p\times k/n$. Then,  
 \be
 {n(k_1,k_2)} \;{ \longrightarrow \atop ^{(p_1,p_2)} }\;   {p_1(k_1,s)}
\oplus  {p_2(-s,k_2)}.\ee
 Especially
interesting cases are those when one of $k_i$ is 1.
 \be
 {n(1,k)} \;{ \longrightarrow \atop ^{(p_1,p_2)} }\;   {p_1(1,s)}
\oplus  {p_2(-s,k)},\;  {\rm if} \; k_1=1, k_2=k.
 \label{transrule1}\ee

In order to check the validity of our method, we check that 
all of examples studied in APS and HKMM, where some
of $k_1=1$ case is considered.  
\begin{enumerate}
    \item    $2l(1,-1) \;{ \longrightarrow \atop ^{(l,l)} }\; l(1,-1)\oplus l(1,-1)$,
    {\rm with} $s=-1$. APS example 5.2
    \item    $2l(1,3) \;{ \longrightarrow \atop ^{(l,l)} }\; l(1,1)\oplus l(1,-3)$,
    {\rm with} $s=1$. APS Ex.5.3
    \item $5(1,3) \;{ \longrightarrow \atop ^{(2,1)} }\; 2(1,1)\oplus \C^2$,
    {\rm with} $s=1$. A generic tachyon condensation. APS Ex.5.4
    \item $n(1,1)\;{ \longrightarrow \atop ^{(p,p)} }
        p(1,0)\oplus p(0,1)$: all charges are on the diagonal $q_1=q_2$
        line, so  $s=0$. This is two copies of $\C^1/\Z_p \times \C$.
    \item $n(1,-1) \;{ \longrightarrow \atop ^{(l,n-l)} }
        l(1,-1)\oplus 'n-l'(1,-1)$: all charges are on the marginal line $q_1+q_2=n$.  $s=-1$.
    \item $n(1,-3)\;{\longrightarrow \atop ^{(j,-3j)} }
        j(1,-\alpha)\oplus \alpha 'n-3j'(\alpha, -3)$, where  $\alpha=[3j/n]+1$.
        Notice $p=(j,-3j)\equiv (j,\alpha n - 3j)$, so that  $s=-\alpha$.
         $\alpha=1$   case is the example 4.3.3 of HKMM.
 \end{enumerate}

Now, what about the  generic case where neither $k_1$ nor $k_2$ is
equal to 1?  We first discuss the non-reducible cases where
$\{lk_i/n\} \neq 0$ for any $l=1,..., n-1$. This is the case if
$k_i$ and $n$ are relatively prime. Then we can choose a new
generator $(1,k)$ such that \be \{ j(1,k)| \; j=1,..., n-1 \}=\{
l(k_1,k_2)| l=1,..., n-1 \}, \ee  because we can find $k$ such
that for any given $l$, $lk_1=j \;{\rm mod} \;n$ and $lk_2=jk
\;{\rm mod} \;n$ for some $j$.  In fact  $k$ is given by \be k
\equiv  {k_2}/{k_1} \;{\rm mod} \; n .\ee Therefore {\it generic
case is isomorphic to  $n(1,k)$  type.}\footnote{So far we proved
this fact in the conformal filed theory level before GSO
projection.
}
 For example, 11(2,3) is identical to 11(1,7) and also to
11(8,1), since $3/2 \equiv 7, \;2/3 \equiv 8  \;{\rm mod} \; 11$.

\begin{figure}[htbp1]
\setlength{\unitlength}{1cm}
\begin{minipage}[c]{6.5 cm}
\hfill
\begin{picture}(6.,6.) \epsfxsize=55mm \centerline{\epsfbox{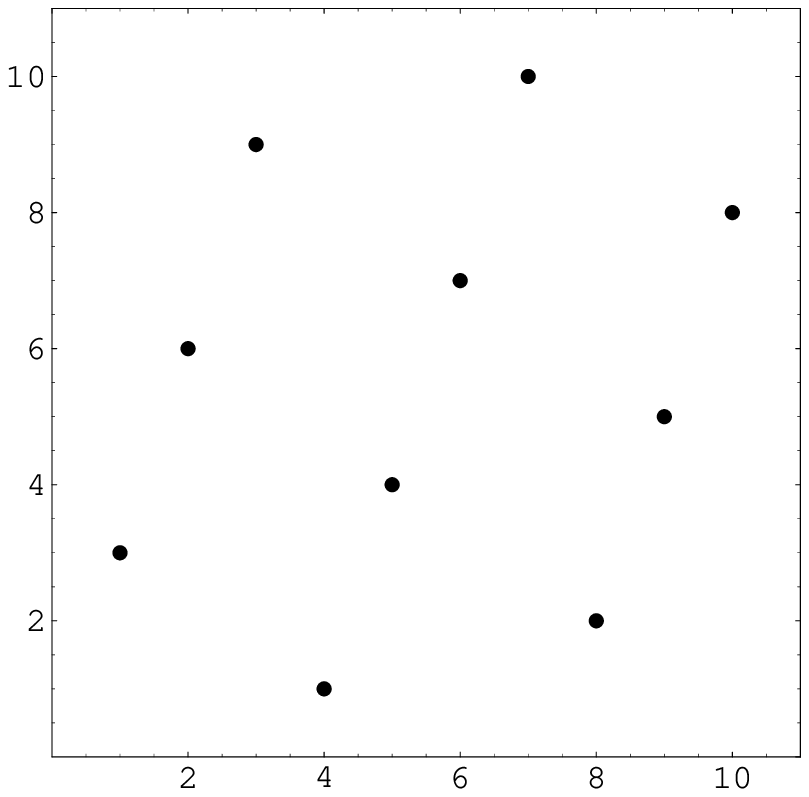}} \end{picture}\par
\end{minipage}
\hspace{1cm}
\begin{minipage}[c]{6.5cm}
\begin{picture}(6,6) \epsfxsize=52mm \centerline{\epsfbox{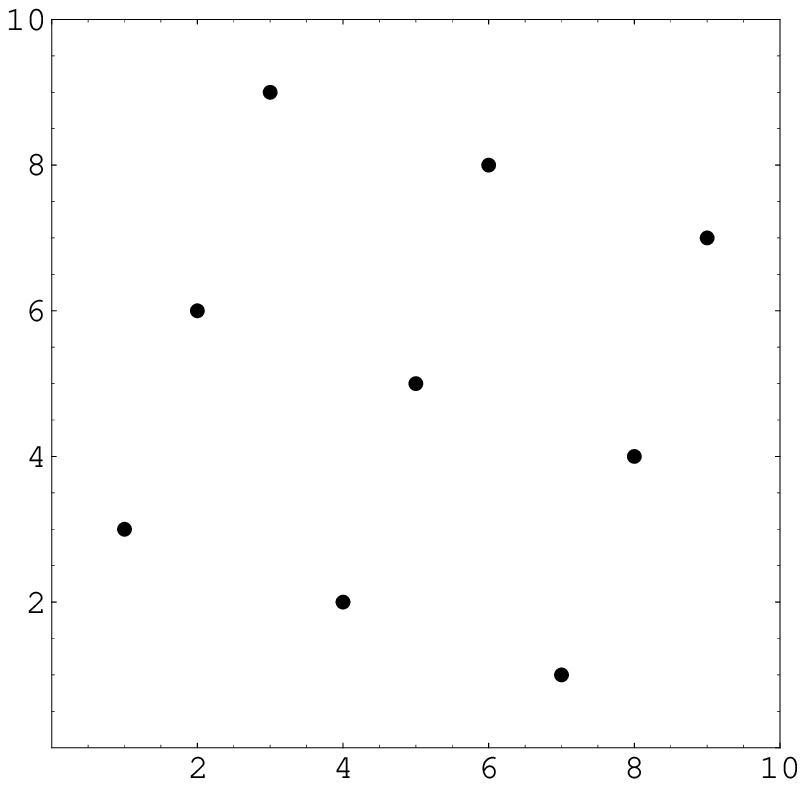} }\end{picture}
\end{minipage}
\caption{\scriptsize  Charges of 11(1,3) (left) and 10(1,3)
(right) in Weight space.}
  \label{cc11-10}
\end{figure}

Sometimes we meet situation where $s=0$, where we need more care.
For example, if we condensate the generator $(1,k)$ itself,
eq.(\ref{transrule1}) predict that \be n(1,k)\to 1(1,0)\oplus
k(0,k).\ee For the first element 1(1,0), it is correct since the
upper triangle does not contain any tachyon operator.  However, for
the second element, this can not be true since we have 
non-trivial operator in the lower triangle. This is clear from 11(1,3) model described in the figure \ref{cc11-10},
where all twisted tachyons coming from chiral primaries are given in the figure \ref{cc11-10}.
 $s=0$ is caused by the fact that
$p$ and $(1,k)$ are parallel. So we need to choose a generator of
the lower triangle other than $(1,k)$. Assuming $k$ and $n$ are
relatively prime, $k$  has   multiplicative inverse modulo $n$,
which we denote by $k^{-1}$. We also introduce
$s'=p\times(k^{-1},1)/n$. Then we have $n(1,k)=n(k^{-1},1)$. Now
the image of the new generator under $T^-_p$ is $(-s',1)$. It is
easy to show that $ks'=s-ap_2$ where $a$ is defined by
$k^{-1}k=na+1$. Therefore $p_2(-s,k) = p_2(-s',1)$ if $s$ is not
0. So we get \be n(1,k) {\longrightarrow \atop ^{(p_1,p_2)}}
p_1(1,s)\oplus p_2(-s',1) . \label{transrule2} \ee The equations
(\ref{transrule1}), (\ref{transrule2}) are the main
formula of this section.
When one of $s,s'$ is 0 and the other is not, we should use the non-zero one.
For example, when the condensing
operator is of the form $j(k^{-1},1)$, $s'=0$ and it is better to
use $p_2(-s,k)$ for the  exactly same reason as we use
$p_2(-s',1)$ when $s=0$.
When  $ss'\neq 0$ two are equivalent in conformal field theory level.
\footnote{For string theory level, two prescriptions are
different if $s$ and $s'$ does not have the same G-parity (even or odd-ness).
we need to use the one that has the same parity  as that of k. This will be discussed
further in later section.}
We give a few examples below.
     If we condensate an operator with $p=j(1,k)$, its band number 
    $G:=[j/n]+[jk/n]=0$ and $s=0$.  However, $s'=j(1,k)\wedge (k^{-1},1)=-aj\neq 0$ unless
$k=1$ ( or, $a=0$). The  transition is
described as \be n(1,k) {\longrightarrow \atop ^{j(1,k)}}
j(1,0)\oplus jk(ja,1) . \ee More explicitly,  for $p=(2,6)$ in
11(1,3), $j=2$, $s=0$, $k=3$, $k^{-1}=4$, $4\cdot 3=11\cdot1+1$
hence $a=1$ and $s'=-2$ so that \be 11(1,3) {\longrightarrow \atop
^{(2,6)}} 2(1,0)\oplus 6(2,1) .
 \ee Notice that $6(2,1)$ contains an operator (0,3)
so that this is a reducible orbifold. Even in the case we start with
irreducible orbifold, we can get reducible orbifold as a result of
tachyon condensation. This happen if and only if there is an
operator sitting on the line which connect (0,0) and the
condensing one, $p$. 
We tabulated all possible tachyon condensation
processes for model $11(1,3)$ and $10(1,3)$  in table \ref{t11.3} and table\ref{t10.3}, respectively.
 
\begin{table}
  \centering
\begin{tabular}{|c|c|c|c|c|} \hline
$j$&$(p_1,p_2)$&$G=[3j/11]$&$ n-(p_1+p_2)$&process\\ \hline \hline
1&(1,3)& 0&7&$11(1,3)\mapsto 1(1, 0)\oplus 3(1,1)$ \\    \hline
2&(2,6)& 0&3&$11(1,3)\mapsto  2(1,0)\oplus 6 (2,1)$ \\ \hline
3&(3,9)& 0&-1&irrelevant process \\    \hline
4&(4,1)&1&6&$11(1,3)\mapsto 4(1,1)\oplus 1(0,1)$\\ \hline
5&(5,4)&1&2&$11(1,3)\mapsto 5(1,1)\oplus4(1,1)$\\ \hline
6&(6,7)&1&-2&irrelevant process\\    \hline
7&(7,10)&1&-6&irrelevant process \\ \hline
8&(8,2)&2&1&$11(1,3)\mapsto 8(1,2)\oplus 2(0,1)$\\ \hline
9&(9,5)&2&-3&irrelevant process \\ \hline
10&(10,8)&2&-7&irrelevant process \\ \hline
\end{tabular}
  \caption{\small \it All possible tachyon condensation process in 11(1,3) model. 
  We should consider only the processes given by relevant operators,
namely those with $n-(p_1+p_2)>0$, otherwise it is a process by an
irrelevant operator which disappears in the infrared limit.}\label{t11.3}
\end{table}
\begin{table}
  \centering
\begin{tabular}{|c|c|c|c|c|}\hline
 $j$ & $(p_1,p_2)$&$G=[3j/10]$&$n-(p_1+p_2)$&process \\ \hline\hline
1&(1,3)&0 &6 &$10(1,3)\mapsto 1(1,0)\oplus 3(0,1)$ \\ \hline
2&(2,6)&0&2&$10(1,3)\mapsto 2(1,0)\oplus 6(0,1)$ \\ \hline
3&(3,9)&0&-2& irrelevant process \\ \hline
4&(4,2)&1&4&$10(1,3)\mapsto 4(1,1)\oplus 2(1,1)$ \\ \hline
5&(5,5)&1&0&$10(1,3)\mapsto 5(1,1)\oplus 5(1,2)$ \\ \hline
6&(6,8)&1&-4&irrelevant process \\ \hline
7&(7,1)&2&2&$10(1,3)\mapsto 7(1,2)\oplus 1(0,1)$ \\ \hline
8&(8,4)&2&-2&irrelevant process \\ \hline 9&(9,7)&2&-6&irrelevant
process \\ \hline
\end{tabular}
  \caption{\small \it All possible localized tachyon condensation in model $10(1,3)$.}\label{t10.3}
\end{table}

\subsection{Equivalence of LG and Toric method in $\C^2/\Z_n$ }
Here we  show the equivalence of our description of tachyon decay in mirror LG model 
with that in toric geometry \cite{martinec} for the case of $\C^2/\Z_n$.   
We will show that the transition in LG picture 
\be n(1,k) {\longrightarrow \atop ^{(p_1,p_2)}}
p_1(1,s) \oplus p_2(-s',1), \ee
with $s=p\wedge(1,k)/n$,
$s'=p\wedge(k^{-1},1)/n$ has corresponding description in toric picture 
\be n(k)
{\longrightarrow \atop ^{(n',-k')}} n'(k')\oplus n''(k''), \ee
where \be n''=kn'-nk' \;{\rm and }\; -k''=cn'-dk'
\label{toricdata}\ee with integer $c,d$ satisfying $cn-dk=1$.
\footnote{If $(c,d)$ is a solution of this equation,
$(c+k'm,d+n'm)$ is also a solution. The result is the $(n''.-k'')
\to (n'',-k''+ n''m)$ which is just an $SL_2Z$ transformation
$\pmatrix{1&0\cr m&1}$ which corresponds to a holomorphic
coordinate transformation of a toric variety.} Notice that it is
assumed that $k,n$ is relatively prime.

The data of weight diagram of LG model can be related to that of
toric geometry by a linear map $U: LG \to  Toric $ and its inverse
$U^{-1}$: \be U= \pmatrix{ 1&0 \cr -k/n&1/n} ,\quad U^{-1}=
\pmatrix{ 1&0 \cr k&n} .\ee  The weight $(p_1,p_2)$ of the
condensing tachyon is related to the corresponding toric data
$n'(k')$ by \be { p_1 \choose p_2}= U^{-1} { n'\choose
-k'}={n'\choose kn'-nk'}, \ee which gives $p_1,p_2$: \be
p_1=n',\quad p_2= kn'-nk', \quad \label{relation}\ee from which
$s$ can calculated in terms of toric data: \be s=p\wedge(1,k)/n =
(n',kn'-nk')\wedge(1,k)/n=k'.\ee Now, since $p_1(1,s)$ is
trivially equal to $n'(k')$, we only need to show the equivalence
of $ p_2(-s',1)$ with $n''(k'')$.   The question is whether
$k''\equiv -s'$ mod $p_2$ or equivalently, \be (cn'-dk') \equiv \;
(p_1-k^{-1}p_2)/n\;\; {\rm mod} \; p_2 \label{eqb}\ee is true or
not. Multiplying both sides by $k$, $(cn'-dk')k \equiv  \;
(kp_1-k^{-1}kp_2)/n {\rm mod} \; p_2$. Using $cn-dk=1$,
$s=(kp_1-p_2)/n$ and $k^{-1}k=1+an$, left hand side is equal to
$k'$ and right hand side is $s-ap_2$. From $s=k'$, we now have
proved eq.(\ref{eqb}). Now $-kk''=ks'$ mod $p_2$ implies
$k''\equiv -s'$ mod $p_2$, provided $k$ and $p_2$ are relatively
prime to each other, completing the proof of our desired result.

Remark: It is interesting to observe that for a general chiral ring element
$q=(j,n\{jk/n\})$, $Uq=(j,k\times q/n)={\tilde T}^+_k(q/n)=(j,-[jk/n])$, which means formally,
$U$ coincide with tachyon condensation mapping for generator condensation.
This fact directly generalizes to the general $(k_1,k_2)$.

\subsection{$\C^3/\Z_n$ }
 
We now describe what happens in $\C^3/\Z_n$ case.
Our method is especially useful in the present case since it applies in this case without any
 difficulty while toric method does not work  here \cite{sarkar}.
When a tachyon with weight vector $(p_1,p_2,p_3)/n$, 
the mirror LG is described by the superpotential 
 \be
[W=u^n_1 +u^n_2+u^n_3 +e^{t\over n}u^{p_1}_1 u^{p_2}_2 u^{p_3}_3 ]//(\Z_n)^2.
\ee
 By considering $u_j\sim 0$ region for  $j=1,2,3$, we get the tachyon maps $T^{(j)}_p$'s \cite{leesin} given by
 \be
{\scriptsize
T^{(1)}_p =
\pmatrix{ 1 & 0 & 0 \cr
          -{p_2 / n} & {p_1 / n} & 0 \cr
            -{p_3 / n} & 0 & {p_1 / n} \cr},\;
T^{(2)}_p=
\pmatrix{ {p_2 / n} &-{p_1 / n} & 0 \cr
          0& 1 & 0 \cr
          0& -{p_3 / n} & {p_2 / n} \cr},\;
T^{(3)}_p =
\pmatrix{ {p_3 / n} &0 & -{p_1 / n} \cr
           0& {p_3 / n} &-{p_2 / n} \cr
           0& 0& 1 \cr} },
\ee 
which play similar role of $T^{\pm}_p$ in $\C^2/Z_n$.
Let $k=(k_1,k_2,k_3)$ be the generator of mother theory. Then  
the the generator of the daughter theories are given by  $k^{(j)}:= T^{(j)}_pk , \; j=1,2,3$. 
Namely the orbifold transition rule is given by 
 \be
 {n(k_1,k_2,k_3)} \;{ \longrightarrow \atop ^{(p_1,p_2,p_3)} }\;   {p_1(k_1,s_{12},s_{13})}
\oplus  {p_2(s_{21},k_2,s_{23})} \oplus {p_3(s_{31},s_{32},k_3)},\ee
where $s_{ji}=p_jk_i-p_ik_j $.  Notice that there exists a simple formula 
\be k^{(j)}_i=k_i\delta_{ji}+s_{ji}.
\ee
This is one of the main result of this paper.  

\section{Conclusion}
In this paper, we determined the  the decay mode of unstable orbifolds by 
working out the generators of orbifold action in daughter theories for  $C^r/Z_n$ $r=1,2,3$. 
We gave a simple method  that works easily and uniformly for all $C^r/Z_n$.
For $C^2/Z_n$, we give  a proof of equivalence of our method  with  toric one. 
Our method trivially reproduced all of 
known cases worked out by brane probe\cite{aps} or toric method \cite{hkmm}. 
For $C^3/Z_n$ cases, the unstable orbifolds decay into  sum of three orbifolds.

Our discussion uses N=2 worldsheet SUSY  essentially. It would be very interesting if we can get the 
the same result without using it. 
 

\vskip .5cm 
\noindent {\bf \large Acknowledgement} \\
I would like to thank   Allan Adams, Keshav DasGupta and Eva Silverstein for discussions.
This work is partially supported by the Korea Research Foundation Grant (KRF-2002-013-D00030) and by KOSEF Grant 1999-2-112-003-5. 
.


\end{document}